# XACs–DyPol: Towards an XACML–based Access Control Model for Dynamic Security Policy

*Tran Khanh Dang* [1]*, *Ha Xuan Son* [2], *Luong Khiem Tran* [3]

[1] *Ho Chi Minh City University of Technology (HCMUT), VNU-HCM, Vietnam*
[2] *Can Tho University of Technology, Can Tho City, Vietnam*
[3] *Aalto University, Greater Helsinki, Finland*
* *Coressponding author's email: khanh@hcmut.edu.vn*

**Abstract:** Authorization or access control is the function of specifying access rights to resources. It plays an essential role in protecting sensitive information and preventing the attackers who are not granted any proper access. The system is based on the security policy to determine an access request allow or deny. However, in recent years, the growing popularity of big data has created a new challenge which the security policy management is facing such as dynamic policy, and update policy in run time. Application of dynamic policy has brought benefits to distributed systems, cloud systems, and social network. To the best of our knowledge, however, there are no previous studies focusing on solving authorization problems in the dynamic policy environments. In this article, we focus on analyzing and classifying when update policy occurs; beside providing solutions for each dynamic policy. The contribution of this paper is two-fold: (1) a novel solution for managing the policy changes even when the access request has been granted by those policies, and (2) providing an XACML-based implementation to empirically evaluate the proposed solution. The experiment section shows the comparison between the XACs–DyPol framework with Balana (an open source framework supporting XACML 3.0). The experiment datasets are XACML 3.0 policy including three samples of real-world policy sets namely Continue-a (266 policies), GEYSERS (7 policies), KMarket (3 policies) and one randomly generated policy set regarding Synthetic-360 (72 policies). According to the comparison results, the XACs–DyPol framework performed better than Balana in terms of all updates in Dynamic Security Policy cases. Specially, our proposed solution outperforms with an order of magnitude when the policy structure includes complex policy sets, policies, and rules or some complicated comparison expression which contains *higher than function* and *less than function*.

**Keywords:** XACML, big data, SMT, dynamic policy, rewriting request, policy evaluation

## 1 Introduction

Big Data is the combination of these three factors: high-volume, high-velocity and high-variety. It requires a particular set of resource arrangements for storing a significant amount of data in few dozen terabytes to many petabytes in a mixture of various data formats. Besides, the large volumes of data may come from distributed sources with high velocity as cloud system, IoTs system, and distributed system. Big Data is beyond the capability of commonly used database systems. Currently, there is a vast of areas that use the achievements of Big Data as a storage system, including business, bioinformatics, agriculture, medicine, finance. Generally, big data is now destroying the barriers between the real and digital worlds. However, one of the enormous problems that can slow down the development of this global wave, or even stop it concerns security and privacy requirements. Indeed, the increasing availability of large and diverse datasets (big data) calls for increased flexibility in access control so to improve the exploitation of the data. Traditional access control mechanisms such as Discretionary Access Control (DAC), Mandatory Access Control (MAC), and Role-Based Access Control (RBAC) schemes are no longer suitable for modeling access control on such a large and dynamic scale as the actors may also change all the time. These changes can occur in attributes, attribute values, and even security policies in the flexible environment.

Attempts to protect the privacy of personal data by including policies from the data subject in an access control system are not new. However, externalizing the authorization is no longer enough to protect access in the eyes of the administrator management. Due to this, many present systems such as cloud systems, distributed systems, as commented on in the literature [1] have complex scenarios. Moreover, the authorization process needs to have dynamicity in its decisions, no more static authorizations (permissions) [2]. Besides, in conventional static policy environments, it is challenging for applications to interwork with new services and devices [3]. In our approach, the dynamic security policy, providing users with a flexible architecture, ensures the effectiveness of operations without compromise to security or privacy. On the one hand, using attribute-based approach is to express the dynamic security policy in terms of security and flexibility, and the dynamic archi- tecture gives dynamic authorization decisions. In the meantime, the dynamic authorization is defined as "the matter of giving (allowing) or taking back (denying) rights and from entities who want to access resources at anytime and anywhere which fulfills all conditions and circumstances meeting policy requirements" [2]. On the other hand, the dynamic security policy is the update policy in the real-time even when the access request has been granted by these policies [4]. Both these concepts must be guaranteed in the present systems. Moreover, Dunlop et al. [3] define the system which can support the dynamic security policy environment if that would involve the following:

- The policy includes a variable containing reference to run-time or location.
- The policy can be altered during the request time.
- Ability to create, update or delete policy during the request time.
- Ability to dismiss assigned policy during the request time.

However, there are still many limitations concerning the dynamic security policy environment issues. Zheng et al. [5] approved that the more a system supports dynamic security policy, the less level of security trust is. Therefore, in what way the balance between two strict requirements of Big Data system, where the policies must guarantee between flexibility and security, can be achieved. To overcome these limitations, Rew-XAC model [6] carries out rewriting the request whenever updating system. The new request $Q^*$ is rewritten based on combining the original request $Q$ and the policy that has the best score computed by a fuzzy function; see [6] for formal definitions. Nguyen et al. [7] extended Rew-XAC model to make it more efficient, flexible and applicable mechanism for data access control in health-care. $Q(DB)$ describes the request $Q$ which want to access the data $DB$, $Q_{P_i}(DB)$ denotes the policy $P_i$ allowing the request $Q$ to access data and $P_{fuz}$ denotes the policy that has the highest fuzzy value. The rewriting request mechanism is defined as follows:

$$Q^*(DB) \subseteq Q(DB) \qquad (1)$$

$$Q^*(DB) = Q_{P_{fuz}}(DB) \cap Q(DB) \qquad (2)$$

However, we identify several fundaments which have not been addressed yet in the two approaches that are unsupported on dynamic update policies. Firstly, the model does not have any support in both updating and deleting conditions not only rules but also policy(set). In the relationship between policies and rules, it is the lack of both combining algorithms and obligation correctly. Finally, they are used only for a particular structure of the policy language in XACML due to the lack of supporting complex Rule and **Policy (Set)** structure.

Rew-SMT better assists in defining XML-standardized policies, which is yielded by XACML v3.0. Besides, the Rew-SMT provides classification mechanisms when the updated policies occur. Nevertheless, performance limitations are still raised by that model, especially when the policies granting access to request have been deleted, and the system is searching for another appropriate policy. If there are not any returned policies {Permit; Deny}, then search for the policies will continue with Indeterminate-valued policies. The system will send authentication requests for missing attributes to the users by using Obligation Service. Based on the results returned, the system will optionally rewrite the query or to return Indeterminate. This approach still places quite a lot of limitations. In the author's paper, the Rew-SMT is believed to be more time-consuming in comparison with conventional approaches if that policy has complex structures and the system consists of numerous attributes. This article is about to expand the Rew-SMT model for better support in the event of complicated structures in policy as well as multiple-attribute systems. Furthermore, our proposed model supports the **Obligation** elements that both Turkmen et al. [8] and Son et al. [4] have not supported yet. Significant contributions to this research are the establishment of Model of Access Control support dynamic security policy (XACs-DyPol). In detail, those contributions are the development of update-classified mechanisms in the event of updating policies, detecting redundancy or conflicting each other between the newly-updated policies and the available ones in the system. Our research also strives for sharpening a simulation (proof-of-concept), supporting Big Data, policy management for both admin and data owners, and aiding dynamic security policy for the system.

The rest of the paper is organized as follows: in Section 2, we provide relevant background materials. Section 3 demonstrates our proposed approach, XACs–DyPol model and our implementation. The following section provides the evaluation. Section 5 reviews prior work and the conclusion and future work are discussed in Section 6.

## 2  Preliminaries

In this section, we introduce the necessary background of XACML policy (Sec. 2.1) and dynamic security policy (Sec. 2.2) in order to understand the remainder of this paper.

### 2.1  XACML Policy

The eXtensible Access Control Markup Language (XACML) is an approved Organization for the Advancement of Structured Information Standards (OASIS) [9]. XACML provides an expressive and extensible syntax in XML for the specification of attribute-based access control policies. The authorization in XACML is based on combining policies possibly specified by independent authorities. The access control policies define conditions to determine which access resources can be granted or denied permission to whom.

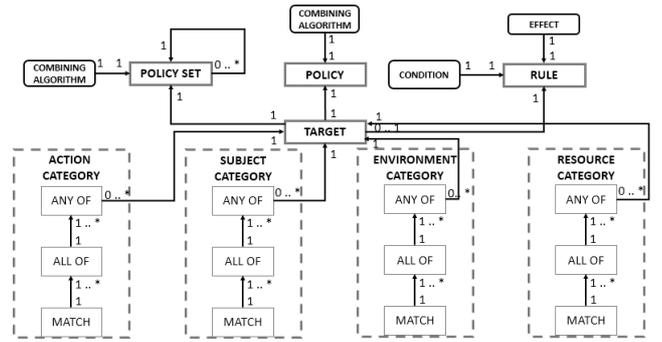

**Fig. 1**: Policy language model [9]

XACML is the de-facto standard for not only an access control policy language but also a request/response language. However, policy evaluation process in XACML is known as a challange and error-prone task [1]. Since XACML was introduced in 2003, a striking number of research works have been done on the policies specification in XACML so far. It focuses on improving evaluation process thanks to optimizing policy structure indirectly.

XACML is receiving a lot of attention because of their effectiveness in authentication problems derived from several application areas. The policy in XACML syntax is based on XML format XML 1.0 specification [10] or based on JSON [11]. In Figure 1, XACML policy has three levels, namely **Policy Set, Policy and Rule** [9]. A **Policy Set** is a root of all policies and includes either the collection of others **Policy Sets** or **Policies** components along with a policy combining algorithm ID and a target. A **Policy** consists of one or more Rules whereas each Rule contains maximum one **Target** and one **Condition** element.

To combine decisions obtained from the evaluation of different applicable policy elements, XACML 3.0 provides some combining algorithms: **permit-overrides** (pov), **deny-overrides** (dov), **deny-unless-permit** (dup), **permit-unless-deny** (pud), **first-applicable** (fa) and **only-one-applicable** (ooa). These algorithms define procedures to evaluate composite policies based on the order of the policy elements and priorities between decisions; see [9] for formal definitions.

As mentioned in figure 1, a **Target** consists of the conjunction of AnyOf component with each AnyOf consists of a disjunction of AllOf components and each AllOf consists of the collaboration of Match. Each Match contains only one particular category to be matched with the request. There are four categories of XACML attributes, i.e., *Subject Category* (e.g., user in the system or the other system etc.); *Action Category* (e.g., read, create, delete, update etc.); *Environment Category* context attributes (e.g., time, date, location etc) and *Resource Category* (e.g., database, server, and so on). **Condition** represents a Boolean expression that refines the applicability of the Rule beyond the predicates implied by its target.

A **Request** contains a set of specific values the on desired access request category, namely *Subject Category* (e.g. ID of user or the other system/service etc.), *Action Category* (e.g. read, create, delete, update etc.), *Environment Category* (e.g. the current time, current date, the temperature, and so on), and *Resource Category* (e.g. object, database, server, etc). Moreover, the request also contains additional information about the external state in *Environment Category* [12].

A **Rule** applicable to a **Request** only if the access request matches both the **Target** and **Condition** elements of the Rule. After that, the decision returns to the value of **Effect** element (Permit/Deny), otherwise the decision is *Not Applicable*. If there is an error during the evaluation process, an *Indeterminate* decision is returned. Additionally, the extended `Indeterminate` value contains the potential `Effect` values which are used to allow for a fine-grained combination of decisions. The possible extended indeterminate values are: [13, 14]

• **Indeterminate Deny** ($IN_D$): an indeterminate from a policy which could have evaluated to deny but not permit, and an `Effect` element of Rule is "Deny".
• **Indeterminate Permit** ($IN_P$): an indeterminate from a policy which could have evaluated to permit but not deny, and an `Effect` element of Rule is "Permit".
• **Indeterminate Deny** ($IN_{PD}$): an indeterminate from a policy which could have an effect on either deny or permit.

### 2.2  dynamic security policy

The dynamic security policy is changed in the run-time. The changes may be due to modifying existing policies, removing policies, adding new policies, adding or removing attributes, updating or upgrading some rules in the existing policies, and so on. The current approaches have started looking at dynamic security policies/rules as opposed to static policies prioritization [2, 13]. Iqbal Hammad et al. can dynamically change the security policies that are appropriate for different applications in SDN by relying on the dynamic security policy [13]. Additionally, Kabbani et al. [2] presented an approach for specifying and enforcing dynamic authorization policies based on situations. Their solution proposes to make a dynamic security policy by modifying authorization rules when the conditions and circumstances are changed.

The access control for dynamic security policy environment should not only support the security requirement, but it should also provide flexible and efficient management of the policy enforced for a large number of users [15]. The users include administrators and data owners who can update the policy anytime and anywhere, or regular users who use system services. Erisa Karafili and Emil Lupu [14] focused on analyzing the various policies and constructing an efficient. The set of policies describe how the data should be shared, used and accessed. Besides, Michelle L. Mazurek et al. [16] focused on solving the assistance for non–expert users who store and share more digital content at home. The approach is that data owners can create specific access policies called `reactive policy creation`. If a user tries to access a resource but lacks sufficient permission, they can use the system to send a request to the resources owner, who can update the policy and allow the access. Both of these approaches used dynamic security policies instead of the static one. This is because new data may be loaded anytime and users may dynamically be assigned or revoked in the multiplex data sharing environment. The requirement changes may be from the data owner or administrator level, such as modifying existing policies, removing policies, adding new rules, and so on. Furthermore, Cullen et al. examined the impact of increasing autonomy on the use of airborne drones in joint operations by collaborative parties [17]. In their dynamic security policy perspectives, doctrine and standards serve as boundaries for a region within which a dynamic security policy decision can be made, and within these constraints, the policy may be modified in response to changing conditions. Therefore, the access control policy must be up-to-date to enforce the right controls.

In figure 2, the dynamic security policy approach is located on the far-right end of the spectrum, where a dynamic security policy and a continuous-time adaptive system are employed for the automation. Security automation strategies of this category allow tailoring and personalization of the security environment for individual users. As a result, they can more effectively reflect security requirements for a given situation [5]. It means that the more dynamism of the policy increases, the lower automatic security is. For that reason, when the systems have any troubles in security policies, due to updated or deleted policy, no higher authority can handle these exceptions. In fact, the system cannot rely on the existing policy, it sends the requirement to administrators or data owner to resolve. However, the policies cannot be managed by a human being given the enormous, heterogeneous amount of data, and data generates with high velocity as cloud system, IoTs system, and distributed system [18]. Consequently, access control that supports dynamically access across the update policies in the dynamic security policy environment is very essential.

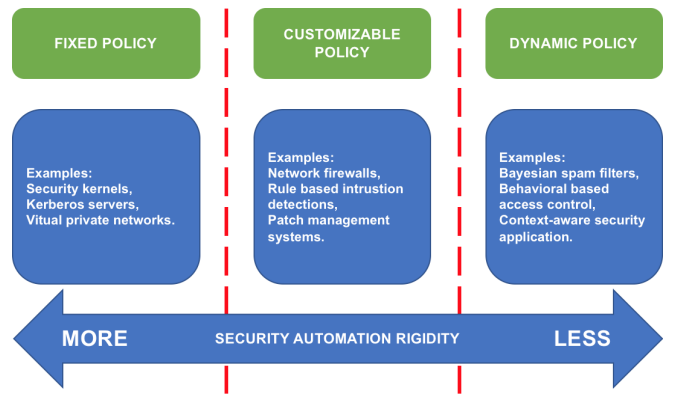

**Fig. 2**: The spectrum of automation approaches [5]

In fact, the policy update is one of the most critical administrative tasks for access control management. Policy based management aims at supporting dynamic adaptability of behavior by changing policy without recoding or stopping the system [19, 20]. This work proposes an access control based XACML with the classification policy update mechanism called XACs–DyPol providing more practical and scalable access control supporting policy updates in the dynamic security policy environment.

### 2.3  Policy Encoding

This section is focused on an overview of the policies included in the system. Policies can be formatted in JSON [11] or XML [9] depending on the defined system. With the initial policies structure, an update is difficult to be detected by system. A transformation of policies is presented to flatten the hierarchical structure of a policy preserving the meaning of the original policy.

The policy is ordered into 3 levels: *PolicySet*, *Policy* and *Rule*. Every element can contain a *Target*. The *PolicySet* can contain the other *PolicySet* or the set of *Policy*. The *Policy* can consist of the different *Policy* or the set of *Rules*. On the other hand, the *PolicySet* and *Policy* have their *Obligations* to fulfill whenever *Response* reaches either *Permit* or *Deny* decision. The following describes the definitions and syntax of all elements.

Obligations *Os* contain one or more obligation(s) *O*. An obligation is an action that takes place after a decision has been reached either `Permit` or `Deny`. It is mapp to XACs–DyPol within the context of policy and policy set according to the following syntax:

$$Os :: = Obligation\_set \qquad (3)$$

$$O :: = \langle OID, FF, AID, DT, Ac \rangle \qquad (4)$$

Where *Obligation_set* is the set of obligation (*O*). *OID* is the ID identifying of obligation, and *FF* stands for the $fulfill$ (On/Off), whose attribute is a boolean function that is used as a key to determine when the obligation must be enforced and must be either `Permit` or `Deny`. *AID* is the attribute ID of the obligation to be carried out. *DT* is the data type and *Ac* is the action to fulfill. If the `Policy` or `Policy set` being evaluated matches the *FF* (On) attribute of its obligations, then the obligations are passed to be enforced otherwise obligations are ignored.

Match *M* contains only one particular category to be matched with the request. There are four categories of XACML attributes, i.e., Subject Category, Action Category, Environment Category, and Resource Category. It is mapped to XACs–DyPol according to the following syntax:

$$M ::= \langle Att_S | Att_{Ac} | Att_E | Att_{Rc} \rangle \quad (5)$$

Where $Att_S$ is the attribute of Subject Category, $Att_{Ac}$ is the attribute of Action Category, $Att_E$ is the attribute of Environment Category, and $Att_{Rc}$ is the attribute of Resource Category,

AllOf *AllOf* contains a conjunctive sequence (one or more) of match element(s). A conjunctive sequence of individual matches the attributes in the request context and the embedded attribute values. It is mapped to XACs–DyPol according to the following syntax:

$$AllOf ::= \overset{\bigwedge}{\phantom{.}} M\_set \quad (6)$$

Where $M\_set$ is the set of match elements $(M)$.

AnyOf *AnyOf* contains a disjunctive sequence (one or more) of AllOf element(s). It is mapped to XACs–DyPol according to the following syntax:

$$AnyOf ::= \overset{\bigvee}{\phantom{.}} AllOf\_set \quad (7)$$

Where $AllOf\_set$ is the set of AllOf elements $(AllOf)$.

Target $T$ is an objective and matching specification for attributes in the context. It contains a conjunctive sequence [zero to many] of AnyOf *AnyOf* elements. If this element is missing, then the target shall match all contexts. It is mapped to XACs–DyPol within the context of rule, policy and policy set according to the following syntax:

$$T ::= \langle \mathbf{null} \mid \overset{\bigwedge}{\phantom{.}} AnyOf\_set \rangle \quad (8)$$

Where $\mathbf{null}$ is the value when missing target element, $AnyOf\_set$ is the set of AnyOf elements $(AnyOf)$.

Condition $C$ is a Boolean function over attributes or functions of attributes. It defines these functions as the expressions which return to a Boolean value, i.e., either true or false. Empty condition is always associated to true. It is mapped to XACs–DyPol according to the following syntax:

$$C ::= \langle \mathbf{true} \mid f^{Bool} Att\_set \rangle \quad (9)$$

Where $\mathbf{true}$ is the value when missing condition, $Att\_set$ is the set of attributes (one or more), $f^{Bool} Att\_set$ is a Boolean function over $Att\_set$.

A rule $R$ is the most primary element of a policy. A rule contains rule conditions, target and rule effect. It is mapped to XACs–DyPol according to the following syntax:

$$R ::= \langle IDR, Cs, T, Re \rangle \quad (10)$$

Where $IDR$ is the identity of rule, $Cs$ is the set of conditions, $T$ is the set of targets, and $Re$ is the rule effect.

A policy $P$ is a single access control policy. It is expressed through a set of rules. A policy contains a set of rules, rule combining algorithm, target and obligations. It is mapped to XACs–DyPol according to the following syntax:

$$P ::= \langle IDP, SR, RCA, T, Os \rangle \quad (11)$$

Where $IDP$ is the identity of policy, $SR$ is the set of rules, $RCA$ is the rule combining algorithm, $T$ is the set of targets, and $Os$ is the set of obligations.

Policy set $PS$ is created by the set of policies $P$ . $PS$ may contain other policy sets, policies or both. It is mapped to XACs–DyPol according to the following syntax:

$$PS ::= \langle IDPS, SP, PCA, T, Os \rangle \quad (12)$$

Where $IDPS$ is the identity of policy set, $SP$ is the set of policies, $PCA$ is the policy combining algorithm, $T$ is the set of targets, and $Os$ is the set of obligations.

### 2.4 Request Encoding

A request $Rq$ is a requirement for access to some resources. It is mapped to XACs–DyPol according to the following syntax:

$$Rq ::= \langle Sr, Ar, Rr, Er \rangle \quad (13)$$

Where $Sr$ is the set of subjects, $Ar$ is the set of actions, $Rr$ is the set of resources and $Er$ is the set of environment information.

### 2.5 Response Encoding

A response $Rs$ is a decision to a request against a based policy. It is mapped to XACs–DyPol according to the following syntax:

$$Rs ::= \langle D, Os \rangle \quad (14)$$

Where $D$ is the decision of the response and $Os$ is the set of obligations to be executed within the response.

### 2.6 Policy Evaluate

The evaluation of the policies starts from the evaluation of Match elements and continues bottom-up until the evaluation of the root of the XACML element, i.e., the evaluation of PolicySet. Each XACML element is denoted by the ranges over the set value when the evaluation process occurs. For example, the evaluating process of Match element ranges over the sets $\{m, nm\}$ which denote *match*, and *no match* while its range is the set $\{t, f, int\}$ standing for *true*, *false* and *indeterminate* when the element is a Condition.

Firstly, we describe the evaluation at target element. There are two different cases namely *match*, and *no →match*. On the one hand, a target $T$ matches a request $Rq$ if the request subject set $Sr$ is the intersect of the target subject set, the request action set $Ar$ is the intersect of the target action set, and the request resource set $Rr$ is the intersect of the target resource set. On the other hand, a target $T$ does not match a request $Rq$ if three sets of requests $(Sr, Ar, Rr)$ are not the intersects of the target ones.

Next, we introduce the evaluation at rule element. The evaluation semantics for a request $Rq$ at rule element $R$ with the set of value as $\{p, d, na\}$ stranding for permit, deny and not applicable. A rule $R$ evaluates a request $Rq$ to Permit if both the target $T$ rule matches the request and the condition $C$ of this rule whose evaluation for request is true and the value of rule effect element $Re$ is Permit. On the one hand, a rule $R$ evaluates a request $Rq$ to Deny if both the target $T$ of this rule matches the request and the condition $C$ of this rule whose evaluation for request is true and the value of rule effect element $Re$ is Deny. On the other hand, the evaluation of this process is Not Applicable if either the target $T$ does not match the request or rule condition $C$ evaluate to False.

Unlike target element and rule element, a policy $P$ evaluates a request $Rq$ to the set of value $\{p, d, in, na\}$ which denote Permit, Deny, Indeterminate, and Not Applicable depend on the value of Rule Combining Algorithm. Figure 3 describe the type of combining algorithm for policy and the ways of combining two rules. The next section shows the evaluation at policy element in the detail.

Similarly, policy, a policy set $PS$ evaluates a request $Rq$ to the set of value $\{p, d, in, na\}$ which stand for Permit, Deny, Indeterminate, and Not Application depend on the value of Policy Combining Algorithm. Figure 3 describes the means of combining algorithm for policy set and the ways of com- bining two policies. The evaluation at policy set element executes similarly policy evaluation.

## 3 Our Proposed Approach

In this section, our proposed model will be put on description and provide the detailed concept of XACs–DyPol and its algorithms.

**Deny-overrides**

$$DS_D^{dov} = DS_D^{p_1} \cup DS_D^{p_2}$$
$$DS_{IN(PD)}^{dov} = \{(DS_{IN(PD)}^{p_1} \cup DS_{IN(PD)}^{p_2}) \cup [DS_{IN(D)}^{p_1} \cap (DS_{IN(P)}^{p_2} \cup DS_P^{p_2})] \cup [DS_{IN(D)}^{p_2} \cap (DS_{IN(P)}^{p_1} \cup DS_P^{p_1})]\} \setminus DS_D^{dov}$$
$$DS_{IN(D)}^{dov} = (DS_{IN(D)}^{p_1} \cup DS_{IN(D)}^{p_2}) \setminus (DS_D^{dov} \cup DS_{IN(PD)}^{dov})$$
$$DS_P^{dov} = (DS_P^{p_1} \cup DS_P^{p_2}) \setminus (DS_D^{dov} \cup DS_{IN(PD)}^{dov} \cup DS_{IN(D)}^{dov})$$
$$DS_{IN(P)}^{dov} = (DS_{IN(P)}^{p_1} \cup DS_{IN(P)}^{p_2}) \setminus (DS_D^{dov} \cup DS_{IN(PD)}^{dov} \cup DS_{IN(D)}^{dov} \cup DS_P^{dov})$$
$$DS_{NA}^{dov} = DS_{NA}^{p_1} \cap DS_{NA}^{p_2}$$

**Permit-overrides**

$$DS_P^{pov} = DS_P^{p_1} \cup DS_P^{p_2}$$
$$DS_{IN(PD)}^{pov} = \{(DS_{IN(PD)}^{p_1} \cup DS_{IN(PD)}^{p_2}) \cup [DS_{IN(P)}^{p_1} \cap (DS_{IN(D)}^{p_2} \cup DS_D^{p_2})] \cup [DS_{IN(P)}^{p_2} \cap (DS_{IN(D)}^{p_1} \cup DS_D^{p_1})]\} \setminus DS_P^{pov}$$
$$DS_{IN(P)}^{pov} = (DS_{IN(P)}^{p_1} \cup DS_{IN(P)}^{p_2}) \setminus (DS_P^{pov} \cup DS_{IN(PD)}^{pov})$$
$$DS_D^{pov} = (DS_D^{p_1} \cup DS_D^{p_2}) \setminus (DS_P^{pov} \cup DS_{IN(PD)}^{pov} \cup DS_{IN(P)}^{pov})$$
$$DS_{IN(D)}^{pov} = (DS_{IN(D)}^{p_1} \cup DS_{IN(D)}^{p_2}) \setminus (DS_P^{pov} \cup DS_{IN(PD)}^{pov} \cup DS_{IN(P)}^{pov} \cup DS_D^{pov})$$
$$DS_{NA}^{pov} = DS_{NA}^{p_1} \cap DS_{NA}^{p_2}$$

**Only-one-applicable**

$$DS_D^{ooa} = (DS_D^{p_1} \cap DS_{NA}^{p_2}) \cup (DS_{NA}^{p_1} \cap DS_D^{p_2})$$
$$DS_P^{ooa} = (DS_P^{p_1} \cap DS_{NA}^{p_2}) \cup (DS_{NA}^{p_1} \cap DS_P^{p_2})$$
$$DS_{IN}^{ooa} = (DS_D^{p_1} \cap DS_P^{p_2}) \cup (DS_P^{p_1} \cap DS_D^{p_2}) \cup (DS_D^{p_1} \cap DS_D^{p_2}) \cup (DS_P^{p_1} \cap DS_P^{p_2}) \cup DS_{IN}^{p_1} \cup DS_{IN}^{p_2}$$
$$DS_{NA}^{ooa} = DS_{NA}^{p_1} \cap DS_{NA}^{p_2}$$

**First-applicable**

$$DS_D^{fa} = DS_D^{p_1} \cup (DS_{NA}^{p_1} \cap DS_D^{p_2})$$
$$DS_P^{fa} = DS_P^{p_1} \cup (DS_{NA}^{p_1} \cap DS_P^{p_2})$$
$$DS_{IN}^{fa} = DS_{IN}^{p_1} \cup (DS_{NA}^{p_1} \cap DS_{IN}^{p_2})$$
$$DS_{NA}^{fa} = DS_{NA}^{p_1} \cap DS_{NA}^{p_2}$$

**Deny-unless-permit**

$$DS_D^{dup} = \mathcal{R} \setminus DS_P^{dup}$$
$$DS_P^{dup} = DS_P^{p_1} \cup DS_P^{p_2}$$
$$DS_{IN}^{dup} = \emptyset$$
$$DS_{NA}^{dup} = \emptyset$$

**Permit-unless-deny**

$$DS_D^{pud} = DS_D^{p_1} \cup DS_D^{p_2}$$
$$DS_P^{pud} = \mathcal{R} \setminus DS_D^{pud}$$
$$DS_{IN}^{pud} = \emptyset$$
$$DS_{NA}^{pud} = \emptyset$$

### 3.1 Scenario

In this section, a sample policy and request presented, used as a running example throughout the paper.

#### 3.1.1 Example Policy:

**The example about Policy $p$**: *When a nurse is in the patient's room within the time interval 8:00 AM - 8:00 PM, she is able to access the personal information of those patients whose age is greater than 16, address is in the south of Vietnam (i.e. Can Tho, Ho Chi Minh, Ca Mau), and disease is the hypertension. One way to model this policy is to represent (the negation of ) these constraints as* **Deny** *rules and then to combine the resulting rules using* **deny overrides** (dov)

$Policy ::= (P, \{R_1, R_2, R_3, R_4, R_5, R_6\}, deny-overrides, \{nurse, read, , patients'srecord\}, \{\})$
$\quad Rule\ 1 ::= (R_1, current-time < 8:00\ AM \lor current-time > 8:00\ PM, \{, , , current-time\}, Deny)$
$\quad Rule\ 2 ::= (R_2, current-place \ne Patient\ room, \{, ,, current-place\}, Deny)$
$\quad Rule\ 3 ::= (R_3, address \not\in \{Can\ Tho,\ Ho\ Chi\ Minh,\ Ca-Mau\}, \{, , , address\}, Deny)$
$\quad Rule\ 4 ::= (R_4, age < 16, \{, , , age\}, Deny)$
$\quad Rule\ 5 ::= (R_5, disease\ Hypertension, \{, , , disease\}, Deny)$
$\quad Rule\ 6 ::= (R_6, , , Permit)$

A single rule is converted into **Attribute** Type (*at*) and **Applicable** **Constraint** (*ac*). The *at* is a variable that stores required attributes from the requesters. The *ac* is a variable that stores conditional constraints to consider the user's requests whether to satisfy the rule/policy/policy set or not. Both *at* and *ac* are linked through logical operations. A policy consists of many rules and is linked together through the Combining Algorithm Rule and so does a policy set, excepting the way of linking, through the Combining Algorithm Policy. The next part is represented the attribute type $at_i$ and applicability constraints $ac_j$ defined from the policy and rules element:

$at_0; at_1; \ldots; at_4 = \emptyset\ (\forall v \not\in Attribute\ Type)$
$Attribute\ Type \in \{current-time, current-place, address, age, disease\}$

$ac_0:$ *"nurse"* $\in$ **subject**
$ac_1:$ *"patients' record"* $\in$ **resource**
$ac_2:$ *"read"* $\in$ **action**
$ac_3: \forall v \in$ **current-time** $v > 8:00PM$
$ac_4: \forall v \in$ **current-time** $v < 8:00AM$
$ac_5: \forall v \in$ **current-place** $v \ne Patient\ room$
$ac_6: \forall v \in$ **address** $v \ne \{Can\ Tho,\ Ho\ Chi\ Minh,\ Ca\ Mau\}$
$ac_7: \forall v \in$ **age** $v \le 16$
$ac_8: \forall v \in$ **disease** $v \ne$ Hypertension

Attribute types and applicable constraints are used to divide the policy space and applicable spaces (AS) are $AS_D, AS_P, AS_{IN}, AS_{NA}$ that denote for applicable space, indeterminate space and not applicable space respectively. We represent the applicability space of a policy element is $AS_{AS}, AS_{IN}$ which means an access request when $req \in AS_{AS}$ iff $req \not\in AS_{AS} \cup AS_{IN}$. The example policy is symbolized as the target of rule $r_i$, $(AS_{AS}, AS_{IN})$ can be presented by $at_i$ and $ac_j$ as follows:

$T_1: ((ac_3 \cup ac_4) \cap \overline{at_0}, \overline{at_0})$

$T_2$: ( $ac_5 \cap \overline{at_1}$, $at_1$ )
$T_3$: ( $ac_6 \cap \underline{at_2}$, $at_2$ )
$T_4$: ( $ac_7 \cap \underline{at_3}$, $at_3$ )
$T_5$: ( $ac_8 \cap at_4$, $at_4$ )
$T_6$: ( $R$, $\varnothing$ )

The decision space $DS$ of rule and policy(set) element is a tuple ( $DS_P$ , $DS_D$, $DS_{IN}$ , $DS_{NA}$ ) such that each element of the tuple denote the value of evaluation i.e. Permit, Deny, Indeterminate, and Not Applicable. The indeterminate decision space $DS_{IN}$ is a triple ( $DS_{IN_P}$, $DS_{IN_D}$, $DS_{IN_{PD}}$ ) representing decisions Indeterminate{Permit}, Indeterminate{Deny} and Indeterminate{Permit Deny}.

$$DS_{IN} = DS_{IN_P} \cup DS_{IN_D} \cup DS_{IN_{PD}}$$

Converting XACML rule into $at$ and $ac$ formulas, we use the notions of **decision space** and **disjoint subjects** as a tuple ( $DS_P^{R_i}$, $DS_D^{R_i}$, $DS_{IN}^{R_i}$, $DS_{NA}^{R_i}$ ) introduced in [4]. We convert the original policy into a decision space of the rules as follow:

$R_1$: ( $\varnothing$, $(ac_3 \cup \underline{ac_4}) \cap \overline{at_0}$, $at_0$, $(ac_3 \cup ac_4) \cap \overline{at_0}$ )
$R_2$: ( $\varnothing$, $ac_5 \cap \overline{at_1}$, $at_1$, $\overline{ac_5} \cap \overline{at_1}$ )
$R_3$: ( $\varnothing$, $ac_6 \cap \underline{at_2}$, $at_2$, $\overline{ac_6} \cap \overline{at_2}$ )
$R_4$: ( $\varnothing$, $ac_7 \cap \underline{at_3}$, $at_3$, $\overline{ac_7} \cap \overline{at_3}$ )
$R_5$: ( $\varnothing$, $ac_8 \cap at_4$, $at_4$, $\overline{ac_8} \cap at_4$ )
$R_6$: ( $R$, $\varnothing$, $\varnothing$, $\varnothing$ )

The next step describes the mechanisms to create a decision space of the overall policy introduced in [4]. The decision spaces induced by deny-overrides define in figure 3. We convert from the original policy to a tuple ( $DS_P$ , $DS_D$, $DS_{IN}$ , $DS_{NA}$ ) with the aim at transforming the decision space into an extension structure presenting the XACML 3.0 policy. The following formulas are the representing of the original policy as the tuple ( $DS_P^P$ , $DS_D^P$, $DS_{IN}^P$ , $DS_{NA}^P$ ) defined from the combination between $at_i$ and $ac_j$ .

$$DS^P : (DS_P^P, DS_D^P, DS_{IN}^P DS_{NA}^P)$$

Where:

$DS_P^P = (ac_0 \cap ac_1 \cap ac_2) \cap \overline{(at_0 \cup at_1 \cup at_2 \cup at_3 \cup at_4)} \cap$
$(((\underline{ac_3 \cup ac_4}) \cap at_0) \cup (ac_5 \cap at_1) \cup (ac_6 \cap at_2) \cup (ac_7 \cap at_3))$
$\cap (ac_8 \cap at_4)$

$DS_D^P = (\underline{ac_0} \cap ac_1 \cap ac_2) \cap (((ac_3 \cup \underline{ac_4}) \cap \overline{at_0}) \cup (ac_5 \cap$
$at_1) \cup (ac_6 \cap at_2) \cup (ac_7 \cap at_3) \cup (ac_8 \cap at_4))$

$DS_{IN}^P = (ac_0 \cap ac_1 \cap ac_2) \cap (at_0 \cup at_1 \cup \underline{at_2} \cup at_3 \cup at_4)$
$\cap ((\underline{(ac_3 \cup ac_4}) \cap at_0) \cup (ac_5 \cap at_1) \cup (ac_6 \cap at_2) \cup (ac_7 \cap \overline{at_3}))$
$\cap (ac_8 \cap at_4)$

$DS_{NA}^P = (ac_0 \cap ac_1 \cap ac_2)$

### 3.1.2    Access Request Example:

**Example request Rq**: *At 8:00 AM, a nurse who is in one patient's room requests to read the personal information of the patient whose age is greater than 50, the address is in Ho Chi Minh, and disease is hypertension* [4].

$Rq$:: = nurse, read, {patients' record, *age* > 50, *address* = Ho Chi Minh, *disease* = Hypertension}, {8:00 AM, patients' room} )

Table 1 summarizes the evaluation process of the **example request** and the **example policy** in the scenario section. The information from the second row to the fourth one presents the response of the comparison in *Subject, Resource, Action*. Thanks to the value in **Result** column we know that the evaluation does not return Not Applicable due to the value of these are **True**. The other values may return Permit, Deny or Indeterminate. In the next five rows, the values in **Result** column are **False**, so $DS_P^P$ is **True**, it means the example request has been granted by the example policy and the system sends the query data to access requester. A response $Rs$ is a decision to a request $Rq$ against a based policy $P$ as followed:
$Rs$:: = ( Permit, {} )

**Table 1** Evaluation Process

| Attribute Type | Policy | Request | Result |
|---|---|---|---|
| Subject | $ac_0$ | nurse | True |
| Resource | $ac_1$ | patient's record | True |
| Action | $ac_2$ | read | True |
| Current Time | $ac_3$ $ac_4$ | 8:00 AM | False |
| Current Place | $ac_5$ | patient's room | False |
| Address | $ac_6$ | Ho Chi Minh | False |
| Age | $ac_7$ | > 50 | False |
| Disease | $ac_8$ | Hypertension | False |

### 3.2    Policy Changes Definition

In the Big Data environment, it is difficult to manage the data sharing; therefore, in order to satisfy the security of data, the system must support the security functions of data owners. One of the most widespread ways of the present-day support is to allow data owners to change their policies to their fullest. However, the difficulty of controlling the change of policy dramatically affects the overall operation of the system. It should be a better mechanism to handle the policy change. This section introduces our approach to address dynamic security policy.

In fact, the dynamic security policy is divided into three types of updates, including insert policy, delete policy and edit policy [3]. Additionally, this article's approach to performance optimiza- tion means focusing only on policy changes that affect the request statements in the system. In the opposite case, updating is implemented independently of the entire system, moreover, increases system processing speed.

#### 3.2.1    Delete policy: the system must find a replacement policy for the deleted policy. If none of the policies satisfies the system, it returns Not Applicable. In case of discovering a replacement policy, the system re-evaluates the policy with the original request before returning the decision. In the above example, if we remove policy $P$ , the request $Rq$ reaches the decision Not Applicable.

#### 3.2.2    Insert policy: In case of a new policy, it does not affect the available request statements in the system that have been assigned decision <AS, IN> (satisfy redundant and conflicting). For the rest, requests with the decision <NA> (not yet satisfied with any policy) evaluated the newly added policy and then returned decision results after being updated to the users. In the above example policy $P$ , when adding any policy, the response $Rs$ is a decision still Permit.

#### 3.2.3    Edit policy: This case consists of three minor cases: insert rule, delete rule and edit rule.

• **Insert rule:** Policy adds a rule to re-evaluate the query to consider if the decision has changed. If the returning result is {Permit, Deny}, the decision value is returned to the users. In case of the return value being Indeterminate, an Obligation has generated and sent the user request to fulfill the requirement. Afterward, the original Request was rewritten before re-evaluating. Considering the above example policy $P$, we assumed that $P$ adds a rule that has the value Department = Heart Center. The request $Rq$ does not meet the requirements, so the return value is Indeterminate. The response $Rs$ returned to the user must be verified by the Department's value as an Obligation:
$Rs$:: = ( {}, { $Os_1$, On, *Department*, String, getString} )
    The decision is likely to change which depends on the value that the users provide. Calling the value provided by the users X, if X = "Heart Center", decision is Permit; if X /= "Heart Center", decision

is Deny. In case the user can not provide the value or exceed the time allowed, then decision = **Indeterminate**.

• *Delete rule:* In the case of **delete rule**, access control system re-evaluates the original query with the updated policy. The decision value is then returned to the users. In the above example policy $P$, we assume rule 4 ($R_4$) to be removed. Then the value remains unchanged (decision is **Permit**).

• *Edit rule:* In the case of **edit rule** access control system re-evaluates the original query with the updated policy. The decision value is then returned to the users. In the above example policy $P$, let's assume to change rule 4 ($R_4$: age > 16). At this time, the decision value comes back to Deny.

### 3.2.4 *Other kinds of dynamic security policy:* 
It is possible to add or delete a policy set $PS$ (similar to adding and removing policies) or to update the value of the **Effect** in the rule $R$ as well as Combining **Algorithm** for both policy and policy set. However, the manager must particularly be careful because each Combining **Algorithm** results in a different purpose. Therefore, carefulness should be taken to avoid accidental changes. Considering the example policy $P$, if the manager changes the Combining **Algorithm** from **Deny-override** to **First Applicable**, then the users only need to satisfy rule 1 ($R_1$) to be allowed. That is the big problem because the **First Applicable** sorts the priority of policies/rules from hight to low. Also, changing the **Effect**s of any rules while the unchanged constraints associated with that rule results in a change in the original rule meaning.

### 3.3 *System Model of Dynamic Security Policy Access Control*

This section firstly presents an overview of our model. In a big data environment (such as distributed systems, cloud systems, IoT systems, and so on), several data generate data streams from difference data owners (a large number of users) with high speed and in large volume. The data can be processed in real-time and integrated into the storage system. One way to protect sensitive permissions data in this environment is to create security policies for these data. One of the most significant issues for the sensitive data stored in the storage systems is to guard against unauthorized access to data, and there are genuine concerns that moving to the dynamic security policies will bring about less, rather than more, stringent security controls. Questions are repeatedly raised around how to applying appropriate dynamic security policy when data owners have any update in policy.

An access control model with these types of properties specifies that accesses are permitted only by a certain subject to a certain object with certain limitations (e.g., object $X$ can be accessed no more than $i$ times simultaneously by user group $Y$). For example, if a user's role is a cashier, he or she cannot be an accountant at the same time when handling a customer's checks. Figure 4 illustrates the system model of big data access control. It shows the interaction between the integrated/loaded data processed (from data owners) and stored in the storage server, our proposed access control system called XACs–DyPol, and critical players including admin, data owners, and users. We assume that our access control scheme is applied to control the authorization of big data which has already been processed and integrated with several data sources. There are two types of policy in our model including the **system policy** which is used to manage the system and store data and the **data policy** which is used to protect the specific resource types. The permissions and functions of the players involving in the system are defined as follows:

1. **Admins:** or "System Administrators" are the entities who have responsibility for the upkeep, configuration, and reliable operation of the system. The admins seek to ensure that the uptime, performance, capacity, and safety of the systems which they manage to meet the needs of the data owners and the users. We assume that admins only manage the system policy and these policies are not conflicted with data policy which managed by data owners.

2. **Data Owners:** or "Data Providers" are the entities who upload their data to the system. They also create the access control policy to regulate how the users gain access the particular resource and what privileges they have over the resource. In our approach, the system has data controllers which are run by the data owners to manage policies (update, insert, delete) and access portions of data. To do so, data controllers permit to access data and data policy. Additionally, we assume that the data policy is only created and updated by data owners.

3. **Users:** or "Data Consumer" is a subject who requests to access (read or write) the big data provided by the data owners in the storage system. Each user is assigned a set of attributes managed by the Attribute Management. These attributes define his/her role and permission.

### 3.4 *Overview of XACs–DyPol dynamic security policy process*

In this section, we give basic system definitions of our proposed access control model called Model of Access Control support dynamic security policy (XACs–DyPol) which was originally proposed in our previous work [4]. However, compared to the work in [4], we support more situations of dynamic security policy that affect the return data for the users. These works mainly focus on how to update policies from the data owners and how to enforce them.

As illustrated in Figure 4, the XACs–DyPol model depicts the relationships between five entity sets including Policy Administrator Points (PAP), Policy Enforcement Points (PEP), Policy Decision Points (PDP), Update Management (UM), and Attribute Management (AM). We describe the relationships and functions of each entity involving in the system as follows:

• **Policy Administrator Points:** The system entity that stores the system policy and data policy. The system policy and data policy are managed by the administrator and data owners respectively. PAP denotes the decision space of the policy as a tuple $\{DS_P^{P_i}, DS_D, DS_{IN}, DS_{NA}\}$ and makes them available to PDP.

• **Policy Enforcement Points:** The system entity that performs access control, by making decision requests and enforcing authorization decisions. PEP receives the access requests from the users and converts these requests into a list of attribute value as the tuple $\{Sr, Ar, Rr, Er\}$. The PEP guards access to a set of resources and ask the PDP for an authorization decision.

• **Policy Decision Points:** The system entity that evaluates applicable policy and renders an authorization decision. The PDP evaluates the process which finds the best policy to grant for each access request. Before the PDP returns a response to the PEP, it converts each response to the tuple $\{DS_P^{P_i} = \{True / False\}, DS_D^{P_i} = \{True / False\}, DS_{IN}^{P_i} = \{True / False\}, DS_{NA}^{P_i} = \{True / False\}\}$. According to these responses, the system could (not) grant a policy to the access request.

• **Update Management:** The system entity that manages and categorizes when policy updates occur. Whenever the policy update happens, the PAP leads both system policy and data policy to handling with the list of at (Attribute Type) and ac (Applicable Constraint). According to the update type (delete, update, insert) of the at and ac, the Update management sends the requirements to PDP (re-evaluating access request) or PAP (rewriting access request) or both. If the policy update does not affect the result returned to the user then the Update Management does not require anything.

• **Attribute Management:** The system entity that acts as a source of attribute values. the AM stores the user attributes and send the attribute value when requested from the PEP. Besides, the AM provides the attribute values whenever a user sends an access request to the system.

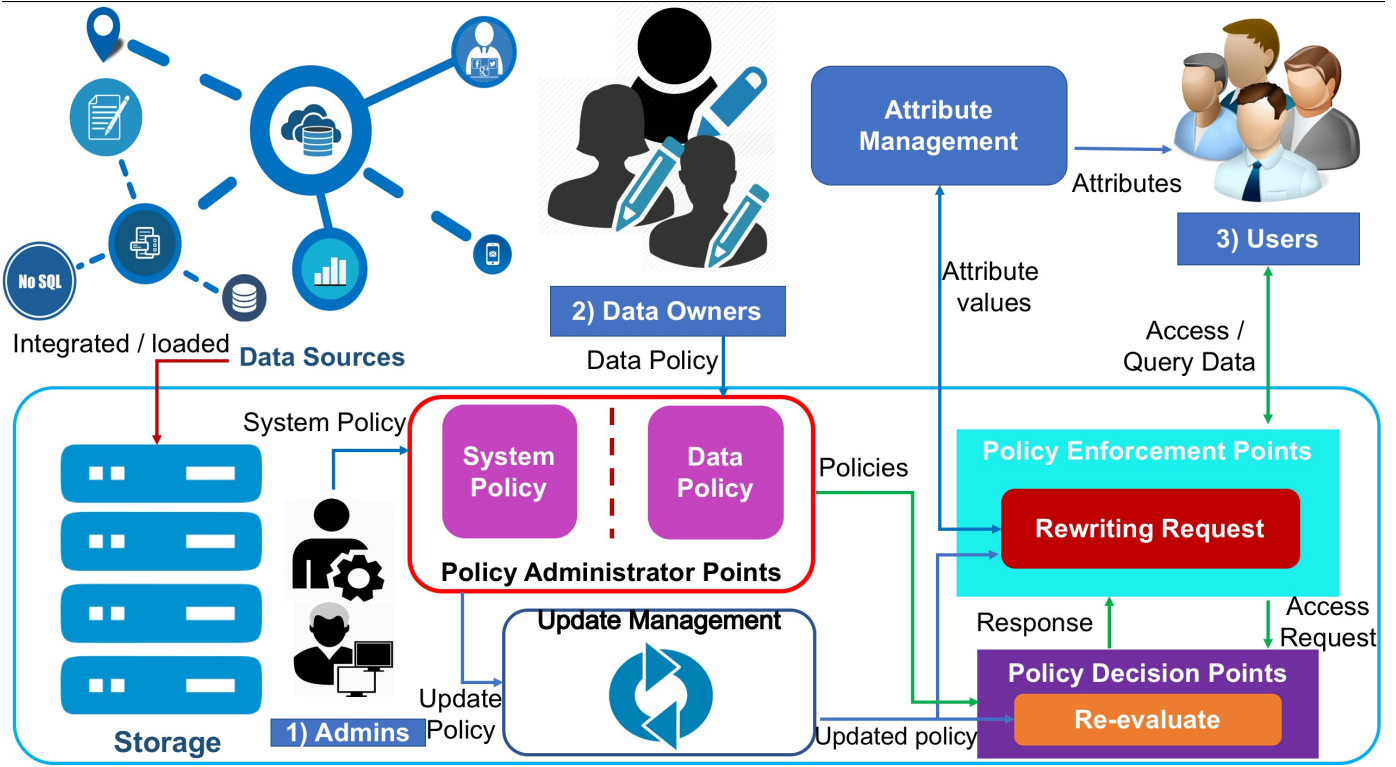

**Fig. 4**: XACs–DyPol: access control model based on XACML for dynamic security policies

## 3.5  XACs–DyPol Algorithm

The section is presented DynamicPolicyEvaluation algorithms. The proposed algorithm is developed based on the extension of the original Rew-XAC [6] by introducing more system parameters based on attribute management, update management and classification update policy.

---

**Algorithm 1** DynamicPolicyEvaluation

---

**Require**: $PAP$ : pap instance; $Q$: request instance
$(decision, non\_applied\_policies, applied\_policy)$ = evaluate($Q$, $PAP$ )
changes = getPolicyChanges($PAP$ , $applied\_policy$)

1: **switch** *changes.type* **do**
2:    **case** *DeletePolicy*
3:       **return**            evaluate($Q$, $PAP$, exclude : $non\_applied\_policies$)
4:    **case** *InsertPolicy*
5:       **return** *decision*
6:       **return**       evaluate($Q$, $PAP$, exclude : $new\_policy$, $non\_applied\_policies$)
7:    **case** *EditPolicy*
8:       **if** *editConstraints* **then**
9:          **return** evaluate($Q$, $PAP$, only : $applied\_policy$*)
10:      **else if** *deleteConstraints* **then**
11:         **return** evaluate($Q$, $PAP$, only : $applied\_policy$*)
12:      **else if** *insertConstraints* **then**
13:         $tempAtt$ = getLackAtt($Q$, $applied\_policy$*)
14:         $v$ = Obligation($tempAtt$)
15:         $Q$* =rewriteRequest($Q$,$v$)
16:         **return**            evaluate($Q$*, $PAP$, only : $applied\_policy$*)
17:      **end if**

---

The Algorithm 1 classifies a policy change from their property (*Delete, Insert* or *Edit policy*). The **Require** of this method includes the PAP instance and the access request. After the policy evaluation process, the result is stored as the tuple (*decision, non_applied_policies, applied_policy*) where *decision* returning to either **Permit** or **Deny**, and *non_applied_ policies* contains the evaluated policy, and *applied_policy* is the policy assigned to the access request. In line 1, *changes* is a parameter storing type of change. If the temp's value is "*Delete*", the method which finds a new policy with the goal of replacing the original policy is deleted. In this process, the *non_applied_policies* policies are ignored (line 2 and line 3). In the type of "*Insert*" the new one is stored in the PAP and the system evaluates the *new_policies* and *non_applied_policies* with the request which does not find the appropriate policy (between line 4 and 6). The case of "*Edit*" is divided into three smaller instances: insert rule, delete rule, and edit rule (line 7). Both delete and edit rule are re-evaluated by the access request with edited policy "*applied_policy*" (from line 8 to 11). For the insert rule, the attributes or evidence will be found if it does not exist in the request (line 12). If not exist, PEP will send an authentication request to **Obligation Service**. Otherwise, *tempAtt* returns null (line 13). Consequently, the Obligation sends these requests to force the users to provide missing attributes and evidence before rewriting and evaluating the access request (line 14 to 16). See [4] for more detail.

## 4  Analysis and Evaluation

### 4.1  Analysis

This section evaluates the performance according to the complexity of each update for both the XACs–DyPol and Balana system. Besides, the complexity of the sample policy sets also affects the performance of the system.

In order to satisfy the "dynamic security policy" for Balana*, we have added the ability to dynamically reload the policy in the

---

**Table 2** Analysis steps for policy update between XACs–DyPol and Balana

|                  | Parse Request | Load Policies | Evaluate Request | Filter Policy | Rewrite Request |
|------------------|:-------------:|:-------------:|:----------------:|:-------------:|:---------------:|
| Delete Policy    | 1             | 2             | 2                | 1             | -               |
| Insert Policy    | 1             | 2             | 1                | -             | -               |
| Delete Condition | 1             | 2             | 2                | 1             | -               |
| Edit Condition   | 1             | 2             | 2                | 1             | 1               |
| Insert Condition | 1             | 2             | 2                | 1             | -               |
| Balana Update    | 1             | 2             | 2                | -             | -               |

database whenever the policy update occurs. Table 2 analyzes the steps taken in each update case. For the Balana system, any updates that occur will reload the policy and then re-value the access request. However, the approach of XACs–DyPol will change for each update scenario. The **Insert Policy** only works with the request which did not found the appropriate policy so "Evaluate Request" again is unnecessary. The three cases: **Delete Policy, Delete Condition**, and **Edit Condition** implement both "Load Policy" and "Evaluate Request" up to two times. In particular, there is one instance **Insert Condition** performs a "Rewrite Request" function in some special cases, as discussed in Section 3.2.3.

### 4.2    Environment and Policy Dataset

In this section, we provide the experiments and results of the performance analysis comparing our scheme to the current approaches. We have implemented the XACs–DyPol framework using Java*.

The system configuration for the experiments is a 64-bit machine with 8GB of RAM and 2.8 GHz Intel Core i5 CPU running macOS High Sierra. The experiment datasets are XACML 3.0 policies. The experiments used three samples of real-world policies (GEYSERS, Continue-a, KMarket) and the randomly generated policy Synthetic-360. These policies are shown in table 3.

• GEYSERS: is a real-life policy taken from GEYSERS project GEYSERS - (Generalised Architecture for Dynamic Infrastructure Services) http://www.geysers.eu/. It just contains the equal function.
• Continue-a: the policy is taken and converted by [21]. It is a policy used to govern the permissions of users such as authors or program committee members in a conference management system [8]. The comparison function in Continue-a is the equal function.
• KMarket: is a real-world policy taken from Balana in 2013. KMarket is the open source and is used to manage authorization in an online trading application. It contains the arithmetic operations whose *equal function* are 58.8% and *higher than function* are 41.2%.
• Synthetic-360: is randomly generated policy using 80% *equality operator* and 20% other complex operators. It also chooses random mixture of all combining algorithms. Synthetic -360 is generated by [21].

**Table 3** Sample real-world policies

| Datasets      | Policy levels | Policy set | Policies | Rules | Attributes |
|---------------|:-------------:|:----------:|:--------:|:-----:|:----------:|
| GEYSER        | 3             | 6          | 7        | 33    | 3          |
| Continue-a    | 6             | 111        | 266      | 298   | 14         |
| KMarket       | 2             | 1          | 3        | 12    | 4          |
| Synthetic-360 | 4             | 31         | 72       | 360   | 10         |

*https://github.com/xuansonha17031991/XACs—DyPol*

Analysis of the complexity between the sample policy sets used in the evaluation of the two XACs–DyPol and Balana systems is based on the following criteria:

• **Policy levels:** The largest number of nested policies in a policy data set.
• **Policies, rules, and rules:** The number of policy sets, policies, and rules are included in any data set.
• **The number of attributes:** The number of all attributes used in the policy data set.
• **Operation:** operations are used to compare and evaluate between policy constraints and request properties.

Table 3 shows the difference between the four sets of policies. It is easy to see that the complexity of Synthetic–360 is higher than the other three sets of policy. GEYSER has the structure and comparison algorithm as simple as it only includes the "equal-string" function. Continue-a produces a more complex structure than KMarket, but the KMarket comparison algorithm has a greater complexity while Continue-a only includes the "equal-string" function. The next review will outline the differences in performance of the two systems on a case-by-case basis.

### 4.3    Testbed

We implement the XACs–DyPol framework and compare the performance with Balana (an open source framework that supports XACML 3.0). This section considers the design of the testbed to compare the performance between our approach and the current one. The current approach is to adopt static policies. The system supports static policies that permanently store policies in the PAP. If any updates occur, those policies will be applied to the next requirements, not the system requirements. Therefore, in order to satisfy the dynamic security policy, the system forces the user to resend access requests. We measured the performance of the XACs–DyPol framework with Balana. Balana is the open source version of XACML 3.0 in a static policy approach. To support dynamic security policy, Balana forces the users to re-submit access requests whenever any policy updates occur. Our measurement scenario is divided into five testbeds.

#### 4.3.1    Insert Policy:
Concerning this testbed, the policies are continually added to the system. This case includes two possibilities: the requests have been granted, and the requests have not found the appropriate policy. The first two capabilities are equally aligned due to the redundant and conflicting requests, so the decision result is returned to the user. In the second possibility, the XACs–DyPol framework only considers requests which have not found the appropriate policy yet and have ignored the previously browsed policies. For Balana, when a new policy is added, Balana must load the whole system as memory functions of the policies that have been browsed because these functions are unavailable.

The figure 5 shows the performance of both approaches. It is clear that, with our approach, the performance is better than Balana. Because the memo function has been granted, it speeds up the computation of the XACs–DyPol framework. In addition, policies with complex operations, however, will not put considerable influence on the performance of the XACs–DyPol, which is affected only by the number of policies, rules, and attributes, as in the case of Continue-a and Synthetic-360. In contrast to XACs–DyPol, as the complexity of the operation substantially increases, it will dramatically affect the performance of Balana. Specifically in the case of KMarket policy, when increasing the number of policies to 4 policies processing time soared while the XACs–DyPol still maintains stable performance.

#### 4.3.2    Delete Policy:
The policies which are chosen randomly would be eliminated in this testbed, which results in two possibilities for access requests. For the first possibility, the removed policy does not grant access to any requests. This capability does not give a significant influence on the system of both approaches. The second possibility is that the policy granting permission for the access request is removed. Our approach would be a policy to replace the deleted policy. The XACs–DyPol framework will not be programmed to search for the previously approved policies; however, it will seek newly added policies or after-updated ones. For Balana, this change affects the system; therefore, the system returns the decision as Not Applicable, and the users have to re-submit the request to access the system. Finding the right policy starts all over again.

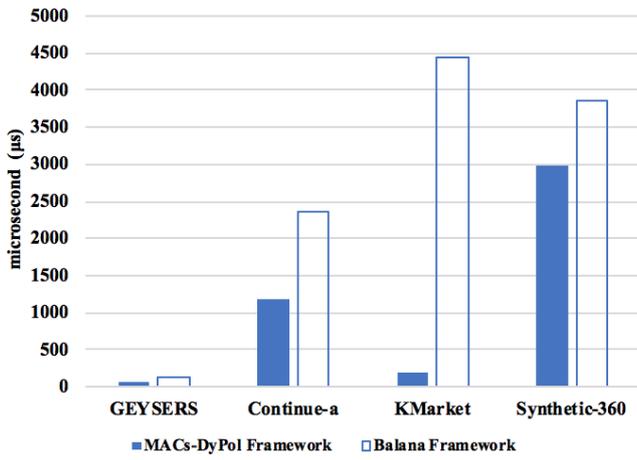

**Fig. 5**: Insert Policy Testbed

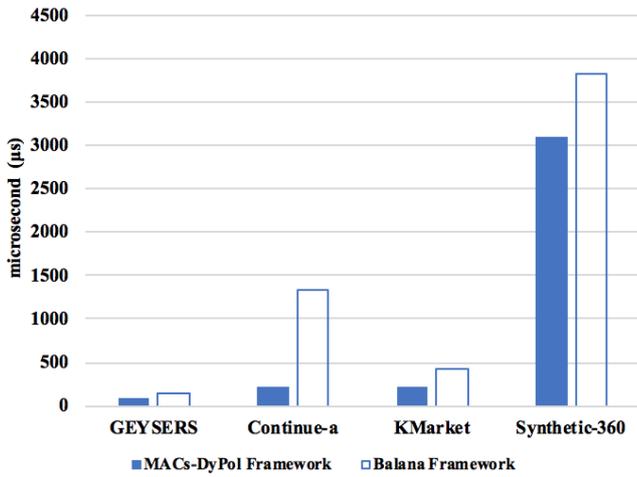

**Fig. 6**: Delete Policy Testbed

Figure 6 shows the performance measurement between the two approaches. In all four sample policy protocols, our approach is more optimistic than Balana because it does not take time to interact with users, but rather to remove previously considered policies and focus on new policies are added, or new policies are updated. Moreover, the highest interval between the two approaches is Continue-a when XACs–DyPol is about four times faster. Processing time for Synthetic–360 is the highest for both systems. In this case, the processing time for Balana's KMarket is considerably reduced with only two policies left.

*4.3.3 Delete Condition:* The policy assigned to the request is changed by deleting any conditions in that policy. With the XACs–DyPol framework approach, the system only re-evaluates before making a decision. However, with Balana's approach, the system notifies the users of updates to policies that affect the outcome of the evaluation process, requests access to be returned, and finds an appropriate policy.

Figure 7 shows the measurement of performance time between two approaches. It is clear that our approach is far more optimum than Balana's approach of focusing on policy change in the XACs–DyPol framework. Moreover, as the complexity grows, so will the processing time of Balana and XACs–DyPol. The highest interval of the two approaches is KMarket, when the processing time of XACs-DyPol is about 12 times faster than Balana. That said, XACs–DyPol is not affected by the complexity of Operation; however, it just depends on the size of the policy file (number of policies and rules). In the case of a small number of policies and complex operations, our approach is far more optimal.

*4.3.4 Edit Condition:* In this testbed, the policy assigned to the request will be adjusted by changing any rules in that policy. That change can be a change of condition, changing the Effect's value. With the XACs–DyPol framework approach, the system only re-evaluates with the policy being updated before issuing the decision. However, with Balana's approach, the system notifies users of updates to policies that affect the outcome of the evaluation process, requests access to be returned, and searches an appropriate policy.

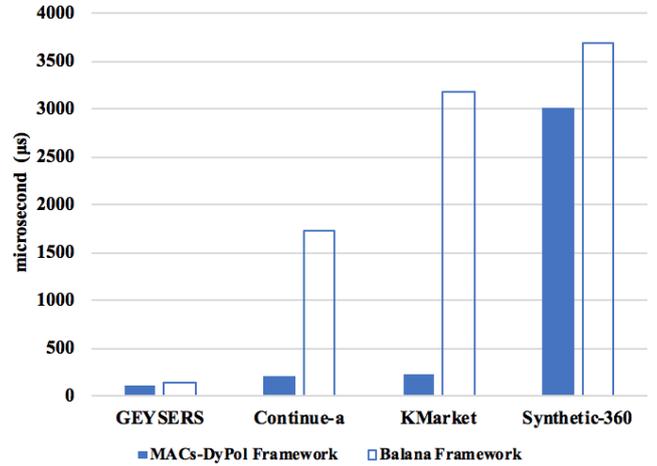

**Fig. 7**: Delete Condition Testbed

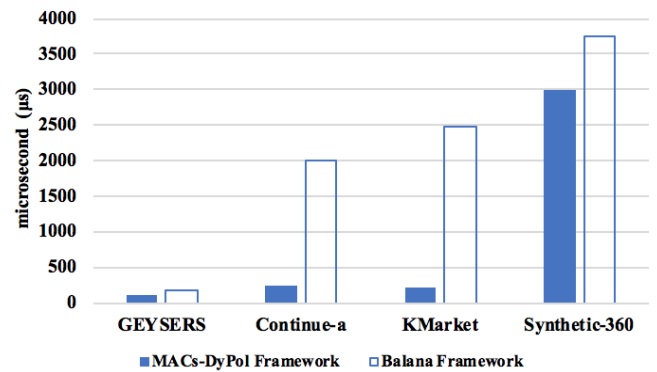

**Fig. 8**: Edit Condition Testbed

Figure 8 shows the measurement of performance time between two approaches. It is clear that our approach is far more optimistic than Balana's approach of focusing on policy change in the XACs–DyPol framework. Besides, as the complexity grows, so will the processing time of Balana and XACs–DyPol. The highest interval of the two approaches is KMarket, when the processing time of XACs-DyPol is about 9 times faster than Balana. That said, XACs–DyPol is not affected by the complexity of Operation; however, it just depends on the size of the policy file (number of policies and rules). In the case of a small number of policies and complex operations, our approach is far more optimal.

*4.3.5 Insert Condition:* In this testbed, the policy assigned to the request is changed by adding any condition in that policy. With the XACs–DyPol framework approach, re-evaluating requests with the changed policy. If the decision is {Permit, Deny} then the system returns to the users. The decision instance is Indeterminate, meaning that the request does not meet the Attribute Type contained in that policy. An Obligation is created later and sent to the fulfill user for this request. PEP then rewrites the original query before re-evaluating it. With Balana, the system notifies users of updates to policies that affect the outcome of the evaluation, requests for access will be resent and find an appropriate policy.

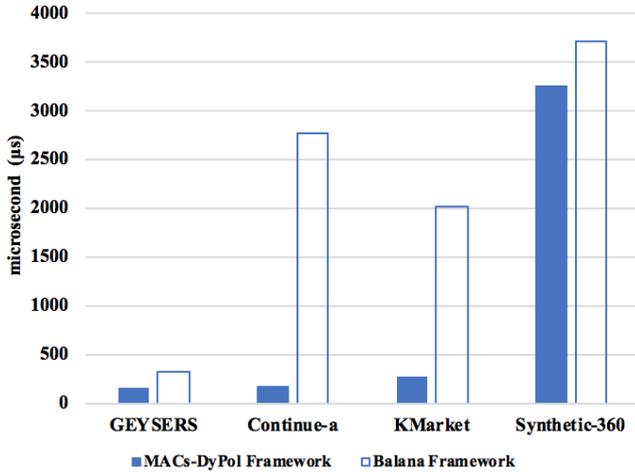

**Fig. 9**: Insert Condition Testbed

Figure 9 shows the execution time of the two approaches, but it is clear that the XACs–DyPol framework is still far time-consuming. Additionally, the processing time for **Synthetic-360** is the highest for both systems. Remarkably, the highest interval between the two approaches is **Continue-a** and KMarket, as the processing time for XACs-DyPol for each case is about 10 and 7 times faster than for Balana, respectively.

### 4.4 Result

The test results show that our XACs–DyPol framework can perform well for all update instances in the dynamic security policy environment. Besides, the execution time is slightly better when compared to Balana on all four template policies. The reason is that this approach focuses only on updated policies and does not spend time on previously approved policies. In addition, the difference between the execution time for all four sample policies has determined that the XACs–DyPol framework has far better results compared to the Balana framework when the policy structure is complicated. The complex policy structure includes multiple properties, multiple nested policy classes, a large number of policies and rules, or the comparison function contains complex arithmetic operations. Furthermore, the difference in execution time for all four policy templates has shown that the XACs–DyPol framework yields far better results than the Balana framework, whose policy field has a complicated operation and not many of policies and rules. In addition, XACs–DyPol does not depend on the complexity of the Operation but the size of the policy file (numbers of policies and rules).

## 5 Related work

### 5.1 Policy Evaluation

Recent work solves the policy evaluation problem directly in different ways: statistic method [22], Decision Diagrams (DDs) [23],

Binary Decision Diagrams (BDDs) [24], Multi-data-types Interval Decision Diagrams (MIDD) [21], Answer Set Programming (ASP) [25]. Marouf et al. [22] based on the statistic of the past access request of the user and its match rules/policies. It means that the approach tries to reorder rules/policies depending on the statistic result. However, the statistic function does not improve the processing time when the access requests are not uniform or the dynamic rules/policies. Liu et al. [23] proposed XEngine, a scheme for efficient XACML policy evaluation. This approach aims at improving the performance of PDP by using decision diagrams in XACML policy evaluation. In particular, they implement numericalization and normalization. *Numericalization* which is the hash function presents all values in policies to integers and stores the results in a hash table. In *normalization*, the major work converts every combining algorithm into **First Applicable** algorithm in order to translate from a policy tree structure to a flat policy structure. Finally, they build a new tree based on numericalized and normalized policies for processing of access request. The limitation of XEngine is that it only solves the part of XACML policy and the numericalization does not support comparison functions. It also ignores obligation processing because it supports only **First Applicable** combining algorithm.

Pina et al. used BDDs theories to build two trees: *Matching Tree* based on the binary search algorithm with an aim at the fast searching of applicable rules, and *Combining Tree* support evaluation of applicable rules in [24]. However, they are only for a special class of the policy evaluation problem due to the lack of complex logical expressions as well as unsupported XACML function. The approach of Ngo et al. [21] is similar to [24] in the spirit. However, they build the intervals from the space of XACML elements matched between an access request and rules/policies. Not only MIDD does improve the evaluation time, but it also provides correctness and completeness of evaluation. However, the approach based on *Decision Diagrams* in the general supports only simple comparison functions (eight *less than, greater than* or *equal*). Additionally, Turkmen et al. [8] claimed that it is very difficult to encode from XACML comparison functions to decision diagrams.

### 5.2 Policy change

Policies represent an externalized logic that can determine the behavior of managed systems [26]. Policy-based management aims at supporting dynamic adaptability of behavior by changing policy without recoding or stopping the system [20]. Laborde et al. [19] introduced how to support the dynamic adaptability of behavior. One of the key motivations of this approach is flexibility and adaptability to the existing infrastructure and the change management. The major contribution of this article has recently implemented self-adaptive authorization frameworks based on XACML that improves the accuracy of a PEP by tracking malicious behaviors. The work uses obligation elements. Our work is different that it depends on updating the original policy definition and carrying out rewriting the access request.

Pallapa et al. [27] proposed an aware scheme for privacy preservation by using user's environment to maintain a model. They implement solution in two cases: for the former case, they account for fine grained rule and for latter case they generate dynamic rules. Nevertheless, the rules are not defined in semantic terms. Additionally, both the rule and the context types are still purposefully based on a set of user activities and states. Ammar et al. [28] approach implicitly updates policy rules based on dynamically inferring a query classification. They do not dynamically update the original policy definitions, but implicitly incorporate context into rule the evaluation.

Son et al. introduced Rew-XAC model in [6], based on XACMLv3. It has been developed to solve the problem in case that requests receive Not Applicable responses from the PDP. The Rew-XAC model carries out the rewriting request by computing a fuzzy function in policy based on an access request. In this paper, the authors concentrate on analyzing the request in case of Not Applicable of PDP response and modifying request based on the conditions of the system's policies. The system allows a request to access data with some restriction, in comparison to the original one. However, they are used only for a special class of the policy structures, which means the policy's structure has only one policy in a policy set and one rule in a policy. Moreover, Rew-XAC theory does not support combining algorithms and multiple deci- sions. The approach of Nguyen et al. [7] is similar to [6] in the content. It focuses on proposing a mechanism integrated with the proposed model to support access control for the data warehouse in the health-care domain. Furthermore, it improves the policy structure which contains more than one rule and multiple decisions. Never- theless, it ignores the obligation processing and does not support combining algorithms. Son et al. [4] extend the Rew-XAC model to be more responsive XACML v3.0 capabilities. Unlike the origi- nal model, Rew-SMT provides a run-time policy update mechanism. However, Rew-SMT does not yet support Obligation element and insert policies update.

### 5.3 Dynamic Authorization

Kabbani et al. [2] presented an approach for specifying and enforc- ing dynamic authorization policies based on situations. The solution of authors proposes to make a policy dynamic by modifying autho- rization rules when the conditions and circumstances are changed. However, the drawback is the rule management complexity when considering many situations and many rules. Additionally, keep- ing loaded rules free of conflict is a big challenge. While, the conflict never occurs in XACs–DyPol because the structure of poli- cies where there is only one **PolicySet** as the root of all of policies and all of others policies are combined by combining **algorithm**. Each combining **algorithm** returns to a sin- gle decision, for example, **Permit**, **Deny**, **Indeterminate** or **Not Applicable** and never returns to more than one decision at the same time.

## 6 Conclusion

We have proposed an access control model for updating security policies in a large data environment by offering dynamic security policy. The policy updates and XACs–DyPol modeling algorithms have been developed to support the safety and efficient development of dynamic security policies in big data. Our proposal reduces the cost of assessing each policy updates. Comparative assessment of the XACs–DyPol framework with the current static policy approach, the results of our assessment show that our approach has been optimized over all four sample policies. Additionally, the XACs–DyPol frame- work can be applied to the actual environment where the systems support changing policies.

For forthcoming work, we are conducting a larger scale of exper- iments with a large number of properties, conditions contained in the policy and higher number of requests. Also, towards the exten- sion of the XACs–DyPol framework, we will integrate additional redundant detection mechanisms in run-time, and then offer hints to users. Moreover, we will consider how to decrease the number of *ac* (applicable constraint) and *at* (attribute type) when encoding policy because of decreasing the complexity of decision spaces. Extending this model for applying to other real-world application contexts like outsourced database services [29] will also be of our great interest in the future.

## 7 References

1 Jaiswal, C., Nath, M., Kumar, V.: 'Location-based security framework for cloud perimeters', *IET Cloud Computing*, 2014, **1**, (3), pp. 56–64
2 Kabbani, B., et al. 'Specification and enforcement of dynamic authorization poli- cies oriented by situations'. In: New Technologies, Mobility and Security (NTMS). (IEEE, 2014. pp. 1–6
3 Dunlop, N., Indulska, J., Raymond, K. 'Dynamic policy model for large evolv- ing enterprises'. In: Enterprise Distributed Object Computing Conference, 2001. EDOC'01. Proceedings. Fifth IEEE International. (IEEE, 2001. pp. 193–197
4 Son, H.X., Dang, T.K., Massacci, F. 'Rew-smt: A new approach for rewriting xacml request with dynamic big data security policies'. In: International Confer- ence on Security, Privacy and Anonymity in Computation, Communication and Storage. (Springer, 2017. pp. 501–515
5 Zheng, Y., Moini, A., Lou, W., Hou, Y.T., Kawamoto, Y.: 'Cognitive security: securing the burgeoning landscape of mobile networks', *IEEE Network*, 2016, **30**, (4), pp. 66–71
6 Son, H.X., Tran, L.K., Dang, T.K., Pham, Y.N. 'Rew-xac: an approach to rewrit- ing request for elastic abac enforcement with dynamic policies'. In: Advanced Computing and Applications (ACOMP), 2016 International Conference on. (IEEE, 2016. pp. 25–31
7 Nguyen, T.T., Pham, N.H.N., Thi, Q.N.T. 'An enhancement of the rew-xac model for workflow data access control in healthcare'. In: International Conference on Future Data and Security Engineering. (Springer, 2016. pp. 251–263
8 Turkmen, F., den Hartog, J., Ranise, S., Zannone, N.: 'Formal analysis of xacml policies using smt', *Computers & Security*, 2017, **66**, pp. 185–203
9 Rissanen, E., et al.: 'extensible access control markup language (xacml) version 3.0', *OASIS standard*, 2013, 22
10 Bray, T., Paoli, J., Sperberg.McQueen, C., Mailer, Y., Yergeau, F.: 'Extensible markup language (xml) 1.0 5th edition, w3c recommendation, november 2008'. (W3C Recommendation,
11 Thi, Q.N.T., Dang, T.K., Van, H.L., Son, H.X. 'Using json to specify privacy preserving-enabled attribute-based access control policies'. In: International Con- ference on Security, Privacy and Anonymity in Computation, Communication and Storage. (Springer, 2017. pp. 561–570
12 Ramli, C.D.P.K., Nielson, H.R., Nielson, F. 'The logic of xacml'. In: International Workshop on Formal Aspects of Component Software. (Springer, 2011. pp. 205– 222
13 Iqbal, H., Ma, J., Mu, Q., Ramaswamy, V., Raymond, G., Vivanco, D., et al. 'Augmenting security of internet-of-things using programmable network-centric approaches: a position paper'. In: Computer Communication and Networks (ICCCN), 2017 26th International Conference on. (IEEE, 2017. pp. 1–6
14 Karafili, E., Lupu, E.C. 'Enabling data sharing in contextual environments: Policy representation and analysis'. In: Proceedings of the 22nd ACM on Symposium on Access Control Models and Technologies. (ACM, 2017. pp. 231–232
15 Fugkeaw, S., Sato, H.: 'Scalable and secure access control policy update for outsourced big data', *Future Generation Computer Systems*, 2018, **79**, pp. 364–373
16 Mazurek, M.L., Klemperer, P.F., Shay, R., Takabi, H., Bauer, L., Cranor, L.F. 'Exploring reactive access control'. In: Proceedings of the SIGCHI Conference on Human Factors in Computing Systems. (ACM, 2011. pp. 2085–2094
17 Cullen, A., Williams, B., Bertino, E., Arunkumar, S., Karafili, E., Lupu, E. 'Mis- sion support for drones: a policy based approach'. In: Proceedings of the 3rd Workshop on Micro Aerial Vehicle Networks, Systems, and Applications. (ACM, 2017. pp. 7–12
18 Outchakoucht, A., Hamza, E.S., Leroy, J.P.: 'Dynamic access control policy based on blockchain and machine learning for the internet of things', *INTERNATIONAL JOURNAL OF ADVANCED COMPUTER SCIENCE AND APPLICATIONS*, 2017, **8**, (7), pp. 417–424
19 Laborde, R., Kabbani, B., Barrère, F., Benzekri, A. 'An adaptive xacmlv3 policy enforcement point'. In: Computer Software and Applications Confer- ence Workshops (COMPSACW), 2014 IEEE 38th International. (IEEE, 2014. pp. 620–625
20 Sloman, M., Lupu, E.: 'Security and management policy specification', *IEEE network*, 2002, **16**, (2), pp. 10–19
21 Ngo, C., Makkes, M.X., Demchenko, Y., de Laat, C. 'Multi-data-types interval decision diagrams for xacml evaluation engine'. In: Privacy, Security and Trust (PST), 2013 Eleventh Annual International Conference on. (IEEE, 2013. pp. 257– 266
22 Marouf, S., Shehab, M., Squicciarini, A., Sundareswaran, S.: 'Adaptive reorder- ing and clustering-based framework for efficient xacml policy evaluation', *IEEE Transactions on Services Computing*, 2011, **4**, (4), pp. 300–313
23 Liu, A.X., Chen, F., Hwang, J., Xie, T. 'Xengine: a fast and scalable xacml pol- icy evaluation engine'. In: ACM SIGMETRICS Performance Evaluation Review. vol. 36. (ACM, 2008. pp. 265–276
24 Pina.Ros, S., Lischka, M., Gómez.Mármol, F. 'Graph-based xacml evaluation'. In: Proceedings of the 17th ACM symposium on Access Control Models and Technologies. (ACM, 2012. pp. 83–92
25 Ramli, C.D.P.K., Nielson, H.R., Nielson, F. 'Xacml 3.0 in answer set program- ming'. In: International Symposium on Logic-Based Program Synthesis and Transformation. (Springer, 2012. pp. 89–105
26 Agrawal, D., Lee, K.W., Lobo, J.: 'Policy-based management of networked computing systems', *IEEE Communications Magazine*, 2005, **43**, (10), pp.69–75
27 Pallapa, G., Di.Francescoy, M., Das, S.K. 'Adaptive and context-aware privacy preservation schemes exploiting user interactions in pervasive environments'. In: World of Wireless, Mobile and Multimedia Networks (WoWMoM), 2012 IEEE International Symposium on a. (IEEE, 2012. pp. 1–6
28 Ammar, N., Malik, Z., Bertino, E., Rezgui, A.: 'Xacml policy evaluation with dynamic context handling', *IEEE Transactions on Knowledge and Data Engineer- ing*, 2015, **27**, (9), pp. 2575–2588
29 Dang, T.K.: 'Ensuring correctness, completeness, and freshness for outsourced tree-indexed data', *Information Resources Management Journal*, 2008, **21**, (1), pp. 59